\documentclass[prl,aps,showpacs,twocolumn,floatfix]{revtex4}
\usepackage{graphicx}
\def\be{\begin{equation}}
\def\ee{\end{equation}}
\begin{document}
\voffset=13mm
\title{Avoided Band Crossing in Locally Periodic Elastic Rods}
\author{R. A. \surname{M\'endez-S\'anchez}}
\altaffiliation{Author to whom correspondence should be addressed}
\email{mendez@fis.unam.mx}
\affiliation{Centro de Ciencias F\'{\i}sicas UNAM, P. O. Box 48-3, 62251
Cuernavaca, Morelos, M\'exico}
\author{A. Morales}
\author{J. Flores}
\altaffiliation{Permanent address: Instituto de F\'{\i}sica,
UNAM, P. O. Box 20-364, 01000 M\'exico, D. F. M\'exico}
\affiliation{Centro de Ciencias F\'{\i}sicas UNAM, P. O. Box 48-3, 62251
Cuernavaca, Morelos, M\'exico}
\begin{abstract}
Avoided band crossings have been studied theoretically and it
has been shown that they can provide a tunning of the metal-insulator
transition. Here
we present an experimental example of an avoided band crossing for a
classical undulatory system: torsional waves in locally periodic rods.
To excite and detect the torsional waves, an
electromagnetic-acoustic transducer for low-frequencies that we have
recently developed, is used.
Calculations performed using the transfer matrix method
agree with the experimental measurements.
In the observed avoided band crossing one level, which is a  border-induced bulk level,
moves from one band to the next.
\end{abstract}
\pacs{43.20.+g, 71.20.-b, 46.40.-f, 62.30.+d}
\maketitle

In a seminal paper, Wigner and von Neumann \cite{WignervonNeumann} studied the
repulsion of the discrete energy levels of quantum systems when an external parameter
$\epsilon$ is varied. For states of the same symmetry they obtained what is called
avoided level crossing \cite{Landau}. In general, however, the spectrum of
a quantum system is not of the discrete type. For instance, periodic systems
show a band spectrum with extended wave  amplitudes according to Bloch's
theorem. If the coupling between these bands is neglected we arrive to one-band
models, such as the Anderson \cite{Anderson} or the Harper models \cite{Peierls,Harper}.
When an external parameter is varied in these systems, some of the bands
move up in energy and others move down,
giving rise to band crossings. At these values of the parameter, however, the coupling
among bands can no longer be neglected. This coupling gives rise to avoided band
crossing (ABC), which is a generalization of the avoided level crossing
mentioned above.

The ABC has been recently studied theoretically in the context of the kicked Harper model
\cite{KetzmerickKruseGeisel} and some
unexpected results were found.
The ABCs can provide a tuning of the metal-insulator transition and changes in the
localization lengths, among other properties.

Although the ABCs have been studied in quantum-mechanical systems, they can also
be obtained for classical undulatory systems such as electromagnetic, elastic or
acoustic systems.
Since we have recently dealt with elastic
waves in rods with locally periodic structures, where wave amplitudes were measured
and a band structure emerges \cite{MoralesFloresGutierrezMendez-Sanchez}, we
are now able to study ABC in elastic vibrations. It is the purpose of this letter to
show experimentally for the first time the existence of an ABC.
Besides their intrinsic interest, the results presented here are relevant for vibration
isolation \cite{Snowdon}
and in the design of very narrow passband filters \cite{Chenetal}.
They could also be of importance in applications of acousto-optic fiber
devices \cite{DiezKakarantzasBirksRussell}
and in the design of many aerospace and marine vehicles \cite{MarcusHoustonPhotiadis}.

\begin{figure}[t]
\includegraphics[width=\columnwidth]{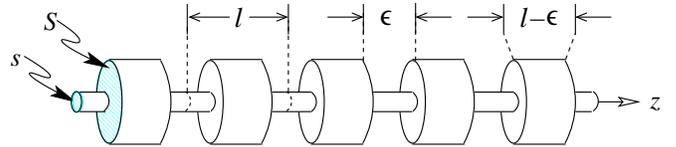}
\caption{Geometry of the rod.  The length of the unit cell is $l$ and
$\epsilon$ is the width of the notch.
Here $S$ and $s$ are the cross section areas of the rod and notches,
respectively.}
\label{rods}
\end{figure}

The rods of finite length $L$ analyzed here consist of
$N$ identical unit cells of length $l$, as shown in
Fig.~\ref{rods}. Each cell is formed by three cylinders, one of length $l-\epsilon$, radius $R$ and
cross section area $S$, and two cylinders of length $\epsilon/2$, radius $r$ and cross section
area $s$.
In this way a rod with $N$ obstacles is formed and $\epsilon$ is the external parameter
to be varied. We assume $R<<L$ and $R<<\lambda=2\pi/k$,
where $\lambda$ is the normal-mode wavelength, so the system behaves indeed as one dimensional.
Torsional waves in this rod then obey the wave equation with a phase velocity
$\sqrt{G/\rho}$, where $G$ is the shear modulus and $\rho$ is the density.
The normal-mode spectrum of the rod is pure point but, as $N$ increases, a band structure emerges
\cite{MoralesFloresGutierrezMendez-Sanchez}.

\begin{figure}[b]
\includegraphics[width=\columnwidth]{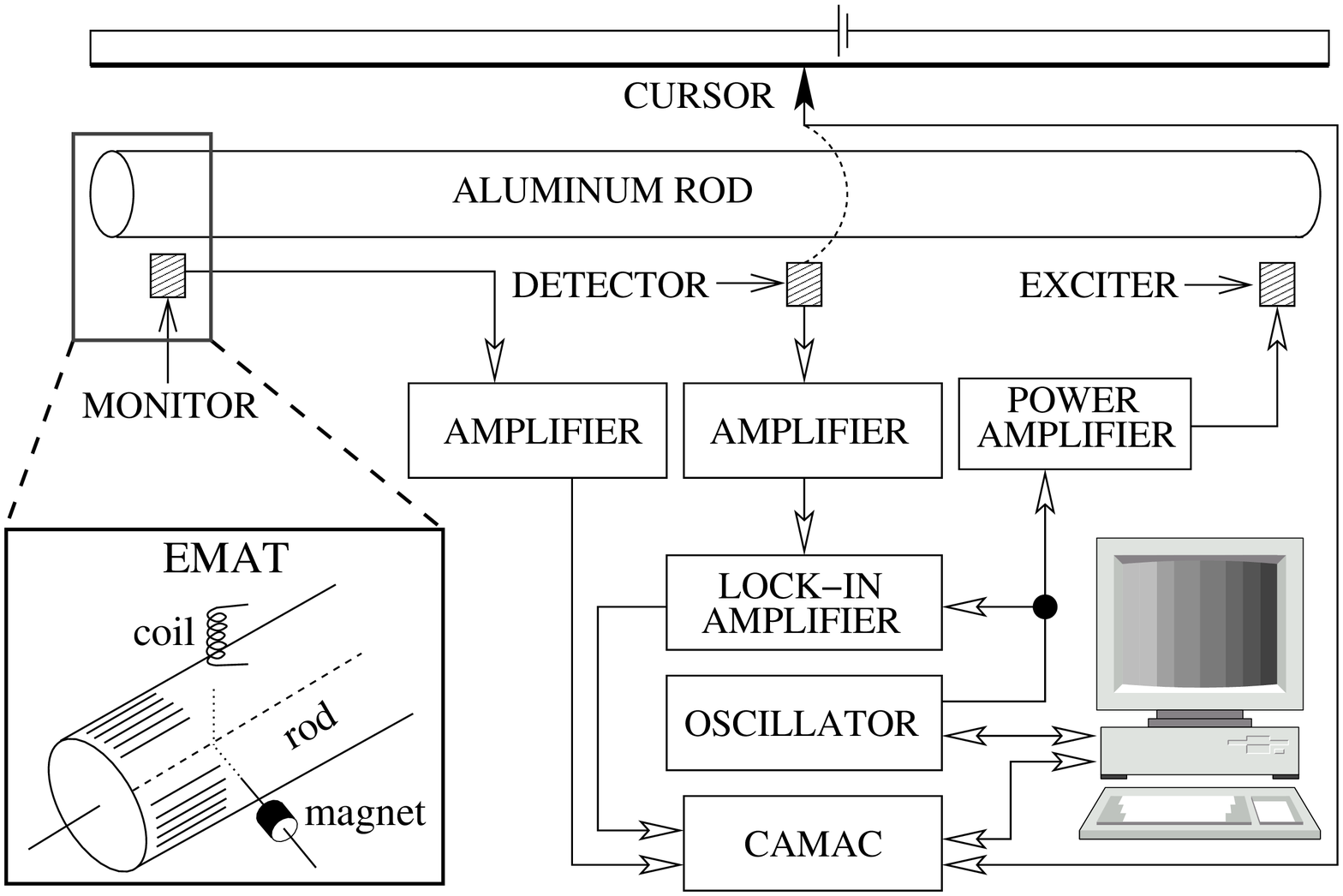}
\caption{Block diagram of the experimental setup.
The EMATs (exciter, detector and monitor) are shown shaded. The dotted line indicates
mechanical contact between the cursor (filled arrow) and the scanning detector.
The schematic diagram
of the EMAT is shown in the left lower corner.}
\label{bloques}
\end{figure}

A diagram of the experimental setup is given in Fig.~\ref{bloques}.
The signal of an oscillator (Stanford Research Systems, Function Generator
DS345) is sent to a power amplifier (Krhon-Hite 7500) and then to the electromagnetic-acoustic
transducer (EMAT) exciter.
This EMAT, as well as the one used as detector,
consists of a coil and a magnet as
shown in the left lower corner of Fig.~\ref{bloques}.
They are configured to excite or detect torsional
waves without any mechanical contact with the rod.
The signal from the EMAT scanning detector is amplified by a
high-impedance amplifier. To monitor the resonances the signal is sent
to the lock-in amplifier (EG\&G PARC 128A). The latter converts
the AC signal to a DC voltage, which is digitized by a CAMAC,
whose output is sent to a PC. The DC voltage is
proportional to the wave amplitude when exciting a normal mode of the rod.
The reference signal for the lock-in was taken
from the oscillator. The position $z$
of the scanning detector along the rod axis is measured by
mechanically coupling it to a cursor in contact
with a nichrome wire, depicted by the bold line of Fig.~\ref{bloques}.
The signal of the voltage divider is then sent to the CAMAC and finally to the PC.
The scanning detector
as well as the cursor are moved with a motor (not shown) controlled by the PC and the
CAMAC.
The normal-mode
wave amplitudes are then scanned along the rod. We should remark, however, that the
resonant frequencies depend on temperature. Instead of controlling the latter, we changed the
frequency of the exciter to follow the resonance. To do this, a third EMAT (denoted
MONITOR in  Fig.~\ref{bloques}) is used to maintain the phase of
the wave constant. The change in frequency is typically of the order of $0.1$\%, which
is therefore the experimental error in the frequency measurements.

The theoretical normal-mode frequencies and their corresponding wave
amplitudes were computed using the transfer matrix
method \cite{MoralesFloresGutierrezMendez-Sanchez}.
In order to do this, we express the wave amplitude
$\psi_i$ in cylinder $i$ as
\be
\psi_i(z)=A_i e^{ik(z-z_{i-1})}+B_i e^{-ik(z-z_{i-1})},
\ee
where $z_{i-1}\le z \le z_i$, $i=1,2,\dots,2N+1$, and impose the following
approximate boundary conditions at the points $z_i$, where the radius changes:
\be
\psi_i |_{z_i} = \psi_{i+1}|_{z_i};  \qquad
\left. s_i^2 \frac{\partial \psi_i}{\partial z}\right |_{z_i} =
\left. s_{i+1}^2 \frac{\partial
\psi_{i+1}}{\partial z}\right |_{z_i}.
\label{torsionalboundaryconditions}
\ee
Here $s_i$  equals either $S$ or $s$ according to the value of $i$. We shall consider
the case of free ends, so Neumann boundary conditions hold at $z_0=0$ or $z_{2N+1}=L$.
More details on the transfer matrix method for torsional waves in rods can be found in
Ref.~\cite{MoralesFloresGutierrezMendez-Sanchez}.

\begin{figure}
\includegraphics[width=\columnwidth]{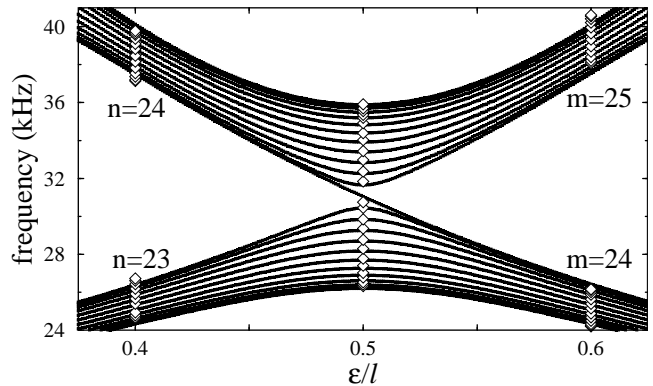}
\caption{Resonant frequencies as a function of $\epsilon/l$ for a
locally periodic aluminum bar with $L=120.0$~cm, $R=6.45\pm0.05$~mm,
$r=3.2\pm 0.05$~mm, and $N=12$.
The experimental values (diamonds) fit rather well with the numerical ones
(continuous lines). The $c$-level corresponds to $n=m=24$ nodes.}
\label{spaghettizoom}
\end{figure}

The spectra of the rods depend on $\epsilon/l$, $\epsilon$ being the external parameter.
When $\epsilon/l \ne 0,1$ a band structure appears and the bands
move up or down as $\epsilon/l$ varies.
The coupling of the bands then leads to avoided band crossings. A typical
ABC is shown in Fig.~\ref{spaghettizoom} for the second and third bands.
The experimental results for
$\epsilon/l = 0.4$, $0.5$ and $0.6$ correspond to the diamonds in the
same figure. A good
agreement is obtained between theory and experiment.
This is remarkable since no fitting parameter was used.
Also, from the same figure it is possible to see that
one level, which we shall call $c$, moves from one band to the other
as the ABC takes place. We will discuss the nature of this level below.

\begin{figure}
\includegraphics[width=\columnwidth]{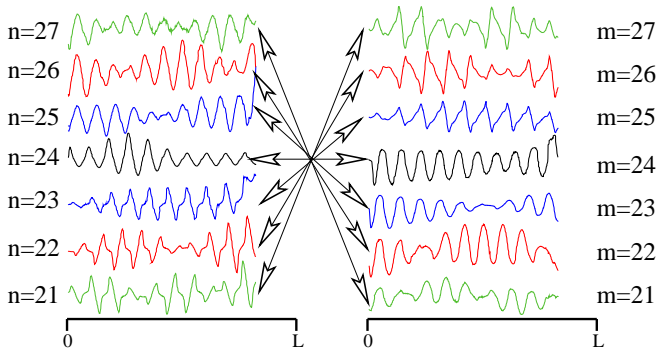}
\caption{Experimental torsional wave amplitudes (in arbitrary units) for an aluminum rod
with $\epsilon/l=0.4$ (left column) and $\epsilon/l=0.6$ (right column).
The amplitudes are not
measured at the right-hand end of the rod since the signal
obtained with the scanning detector is distorted by the exciter.}
\label{amplitudesexp}
\end{figure}

\begin{figure}
\includegraphics[width=\columnwidth]{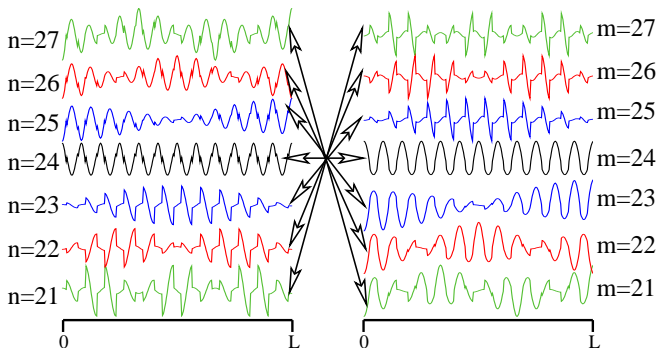}
\caption{Torsional wave amplitudes belonging to the second and third bands
obtained with the transfer matrix method.
The normal modes for $\epsilon/l=0.4, 0.6$ (left and right columns) are labeled by
$n=21,\dots,27$ and $m=21,\dots,27$, respectively.
}
\label{amplitudesteo}
\end{figure}

In Fig.~\ref{amplitudesexp} we present some experimental wave
amplitudes with number of nodes $n=21,\dots,27$ for $\epsilon/l=0.4$ and
$m=21,\dots,27$ for $\epsilon/l=0.6$; they belong to the second and
third bands, respectively.
These are the wave amplitudes at the left- and right-hand side of
the ABC, which occurs
at $\epsilon/l=0.5$, and correspond to the normal mode frequencies
(diamonds) of Fig.~\ref{spaghettizoom}.
The results calculated with the transfer matrix method are given in Fig.~\ref{amplitudesteo}.
A nice agreement with the experimental wave amplitudes is obtained once we take
into account the fact that at the position of the notches
the detector lies farther away than when it is located above the cylinders of radius $R$.
In our case, the experimental signal is reduced by a factor of the order of 8, so the
theoretical values at the notches were divided by this factor.

\begin{figure}
\includegraphics[width=\columnwidth]{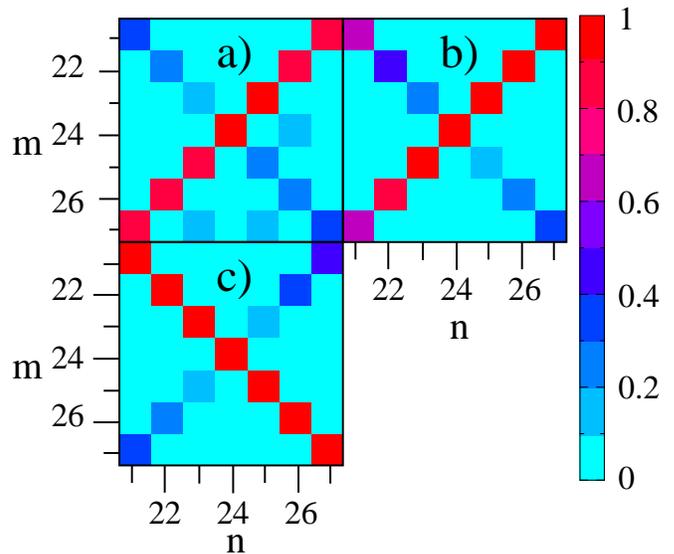}
\caption{Overlap $\mathcal{O}_{nm}$ between wave amplitudes at $\epsilon/l=0.4$ and
$\epsilon/l=0.6$; a) experimental results, b) numerical
results. In c) is shown the overlap between theoretical wave amplitudes at
$\epsilon/l=0.35$ and $\epsilon/l=0.4$.}
\label{correlation}
\end{figure}

As observed in these figures each wave amplitude in the second band is
interchanged with one
belonging to the third band located symmetrically in frequency with
respect to the $c$-level, which corresponds to $n=24$ in
Figs.~\ref{amplitudesexp} and~\ref{amplitudesteo}. For example, the
wave amplitude for $\epsilon/l=0.4$ and $n=22$ is very similar to that
with $m=26$ for $\epsilon/l=0.6$. They are, however, different
because they do not have the same number of nodes.
But the differences are negligible since they occur at points where
the envelope of the
wave amplitudes is negligible and a node bifurcates into three nodes.

To quantify  how much the experimental wave amplitudes of the left-hand side of
Fig.~\ref{amplitudesexp}
mix with those at the right-hand side of the same figure,
we plot in Fig.~\ref{correlation}a the absolute value of the overlap
\be
\mathcal{O}_{nm}=\left | \int_{0}^{L} dz \psi^n(z;\epsilon/l=0.4)  \psi^m(z;\epsilon/l=0.6)
\right |
\ee
between the amplitudes $\psi^n(\epsilon/l=0.4)$ and $\psi^m(\epsilon/l=0.6)$ with
$n,m=21,\dots,27$. The overlap is larger for wave functions that
are symmetrical with respect to the $c$-level while it is negligible for other levels within the
two bands involved in the ABC. This yields an anti-diagonal overlap matrix.
The theoretical overlaps corresponding to the same wave amplitudes are given in
Fig.~\ref{correlation}b.
Thus, experiment and theory agree very well and the overlap between theoretical
and experimental wavefunctions is always larger than $0.84$.
These anti-diagonal correlations only appear when an ABC is present.
To show this, in Fig.~\ref{correlation}c we present a numerical calculation
of $\mathcal{O}_{nm}$ for wave amplitudes at $\epsilon/l=0.4$ and $\epsilon/l=0.45$,
since in this range of $\epsilon/l$ no ABC is present. The overlap
matrix is now concentrated along
the diagonal since the wave amplitudes remain in its position within the band.

\begin{figure}[b]
\includegraphics[width=\columnwidth]{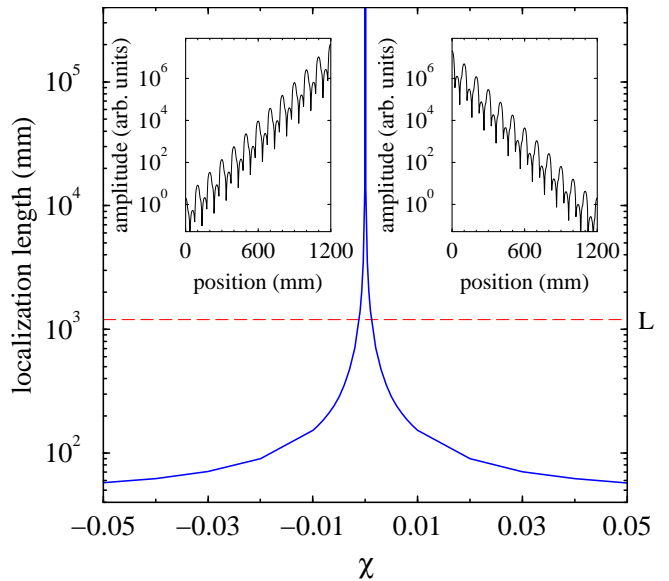}
\caption{Localization length of wavefunction $n=24$ as a function of $\chi$ for
$\epsilon/l=0.4$.
The dashed line indicates the length $L$ of the rod. The insets at the left
and right upper corners show the wave amplitudes for $\chi=-0.03$
and $\chi=0.03$, respectively.}
\label{localization}
\end{figure}

So what is the nature of the $c$-level? It cannot be a Bloch-type state
since at the ABC this level moves from one band to a neighboring one.
Its existence must therefore be due to the fact that the system is finite
so level $c$ is really a border state. To corroborate this we made the
following transformation: $z_i \rightarrow z_i + \chi \epsilon /2$,
with $i=1,\dots,2N$, which alters the two cylinders at the extremes,
the system remaining otherwise locally periodic. The $c$-level is then
localized. The localization length as a function of $\chi$ is given in
Fig.~\ref{localization}. We see that when $\chi=0$, the case
corresponding to Figs.~\ref{amplitudesexp} and~\ref{amplitudesteo},
the localization length is much larger than $L$, so the $c$-level appears
to be an extended state, such as the one labeled $n=m=24$. The
$c$-level is actually a border-induced bulk state \cite{OlguinBaquero}.

In conclusion, we have measured for the first time an avoided band
crossing, in this case for torsional vibrations of elastic rods. Since
we deal with a finite rod a level shifts from a given band to the next one.
This state is an example of a border-induced bulk mode, as predicted
in Ref.~\cite{OlguinBaquero}.
We have obtained a similar behavior for compressional waves and it
is also expected for light and
microwaves \cite{Dembowskietal,KuhlStockmann}.

We thank A. D\'{\i}az-de-Anda and A. Pimentel for their help in
the experimental measurements. We also thank L.  Guti\'errez,  G. B\'aez and J. Maytorena
for useful discussions and L. Moch\'an for a careful reading of the manuscript.
This work was supported by DGAPA-UNAM project IN104400.
\bibliographystyle{prlsty}
\bibliography{thesis,paperdef,paper,newpaper,book}

\end{document}